%% file: 20pcsample.paper1.v5.tex
\documentclass[12pt,apj]{emulateapj}

\shorttitle{A Volume-limited Sample of 63 M7--M9.5 Dwarfs I}
\shortauthors{Reiners \& Basri}

\begin{document}


\title{A Volume-limited Sample of 63 M7--M9.5 Dwarfs\\
  I. Space Motion, Kinematic Age, and Lithium}


\author{A. Reiners\altaffilmark{*}}
\affil{Institut f\"ur Astrophysik, Georg-August-Universit\"at,
  Friedrich-Hund-Platz 1, 37077 G\"ottingen, Germany}
\email{Ansgar.Reiners@phys.uni-goettingen.de}
\and
\author{G. Basri}
\affil{Astronomy Department, University of California, Berkeley, CA
  94720 }
\email{basri@berkeley.edu}


\altaffiltext{*}{Emmy Noether Fellow}


\begin{abstract}
  In a volume-limited sample of 63 ultracool dwarfs of spectral type
  M7--M9.5, we have obtained high-resolution spectroscopy with UVES at
  the Very Large Telescope and HIRES at Keck Observatory. In this
  first paper we introduce our volume-complete sample from DENIS and
  2MASS targets, and we derive radial velocities and space motion.
  Kinematics of our sample are consistent with the stars being
  predominantly members of the young disk. The kinematic age of the
  sample is 3.1\,Gyr. We find that six of our targets show strong Li
  lines implying that they are brown dwarfs younger than several
  hundred million years. Five of the young brown dwarfs were
  unrecognized before. Comparing the fraction of Li detections to
  later spectral types, we see a hint of an unexpected local maximum
  of this fraction at spectral type M9. It is not yet clear whether
  this maximum is due to insufficient statistics, or to a combination
  of physical effects including spectral appearance of young brown
  dwarfs, Li line formation, and the star formation rate at low
  masses.
\end{abstract}





\section{Introduction}

Ultracool dwarfs -- main-sequence stars and brown dwarfs of spectral
class M7--M9.5 -- are of particular importance for our understanding
of Galactic physics for a number of reasons. First, M dwarfs are the
most numerous stellar species and represent an important but poorly
understood contribution to our Galaxy. Our closest neighorhood is
populated by many M dwarfs of which a large fraction remains
undetected. Second, at least three physically important regimes exist
across the M dwarf spectral classes, just at or close to the
temperature range of ultracool dwarfs. These are 1) the star--brown
dwarf boundary; 2) the threshold between partially and fully
convective stars with potential implications on the magnetic dynamo
and activity; and 3) the transition from ionized to predominantly
neutral stellar atmospheres. We discuss each of these in turn below.

Stars and brown dwarfs differ by the force that balances their
gravitational pressure. In a star, central temperatures and pressure
are high enough to start sufficient hydrogen burning that stabilizes
the star. In a brown dwarf, electron degeneracy sets in before
hydrogen burning so that a brown dwarf cannot reach a stable state but
becomes cooler and cooler as it ages. Objects less massive than $M
\approx 0.07$\,M$_\odot$ are brown dwarfs, more massive objects are
stars \citep{Chabrier00}. The most massive and youngest brown dwarfs
can have spectral types as early as mid-M but enter the L dwarf regime
after about 1\,Gyr \citep{Burrows97, Baraffe98}.

In main-sequence stars, the threshold between partially and fully
convective stars occurs at about $M \approx 0.3$\,M$_\odot$
\citep{Siess00}. Partially convective stars harbor a tachocline, the
boundary layer between the outer convective envelope and the inner
radiative core. The tachocline is believed to be the place where the
large-scale solar, and by analogy also the sun-like stellar dynamo is
situated \citep{Ossendrijver03}. This type of dynamo cannot operate in
fully convective stars. However, no change in magnetic activity or the
magnetic field strengths is observed among M dwarfs \citep[see,
e.g.,][]{Mohanty03, Reiners07}. On the other hand, the timescale for
rotational braking and the topology of magnetic fields seems to change
at this threshold \citep{Reiners08, Reiners09}.

The third important threshold in the M dwarf regime, the transition
from ionized to partially neutral atmospheres, occurs among the
ultracool dwarfs of spectral type about M9, i.e., at temperatures of
about 2400\,K \citep{Meyer99, Fleming00, Mohanty02}.  There is strong
observational evidence that around spectral type M9, normalized
H$\alpha$ emission gradually decreases with spectral type, i.e. with
temperature \citep[e.g.][]{Mohanty03}, and that rotational braking
becomes very inefficient at this point \citep{Reiners08}.

This is the first of two papers in which we report on high resolution
spectroscopy in a volume-complete sample of ultracool M dwarfs. In
this one, we introduce the sample and search for young brown dwarfs
via the lithium test. We derive radial velocities and report space
velocities, from which we derive the kinematic age. In a second paper,
we will investigate the properties related to activity, which are
rotation, H$\alpha$ emission, and magnetic flux.

\section{Sample selection and observations}

\subsection{The sample}

We constructed a sample of known M dwarfs of spectral type M7--M9.5
that is almost volume-complete ($d < 20$\,pc). The targets are taken
from several catalogues and discoveries from the DENIS and the 2MASS
surveys.

\citet{Cruz07} present a volume-limited sample from 2MASS that
contains objects in the spectral range M7--L8. Distances are derived
from spectro-photometry with $M_J$ estimated from the
spectral-type/$M_J$ calibration from \citet{Cruz03}. This $M_J$ is
combined with photometry from the 2MASS Second Incremental Data
Release PSC to yield $M_{K_s}$ and spectrophotometric distances.
Uncertainties in distance are dominated by the uncertainties in
spectral type \citep[see][]{Cruz03}. This sample covers about 40\% of
the sky.  The sample taken from \citet{Cruz07} was carefully
constructed to yield a sample that is complete and volume-limited with
$d < 20$\,pc in the spectral range we consider. Nevertheless,
\citet{Cruz07} note that the sample may miss several objects at M7 due
to the $(J - K_s) > 1.0$ selection criterion.

\citet{Crifo05} and \citet{PhanBaoBessel} present two parts of a
sample from DENIS that covers most of the late-M dwarfs up to spectral
type M8.5. Two later objects were discovered in \citet{Delfosse01} and
\citet{PhanBao06}, which make this sample almost complete up to M9.5
while covering $\sim 13$\% of the sky.

For 15 targets, we found parallax measurements in the literature,
which we use instead of the spectrophotometric distances where
available. One target, 2MASS J0041353-562112, was recently reported to
show evidence of accretion, which implies that it is very young
\citep[$\sim 10$\,Myr;][]{RE09}. At this age, the object is much
brighter affecting the spectrophotometric distance, and we use the
distance reported by \citet{RE09}. Taking into account parallax and
age information, three of the 63 targets are found to be located at a
distance $d > 20$\,pc. We did not exclude these targets from the
following investigation.

\begin{figure}
  \plotone{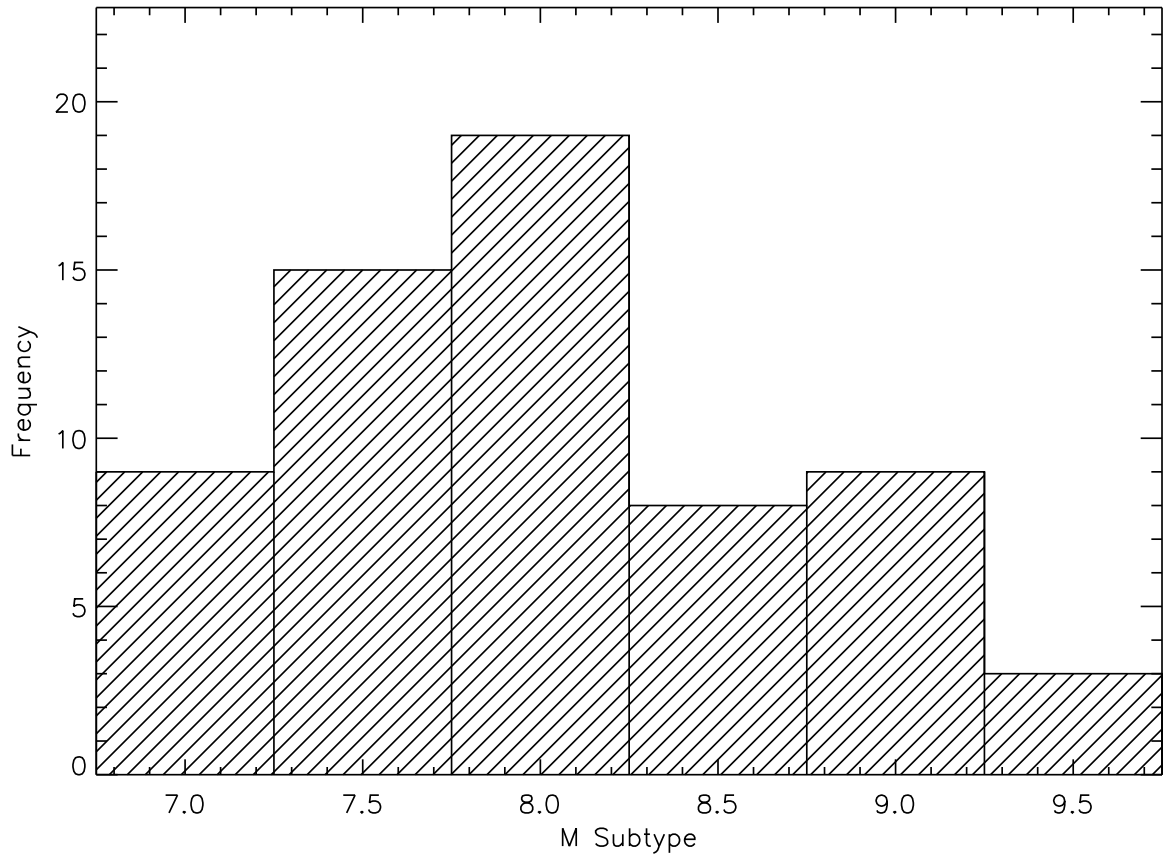}
  \caption{\label{fig:SampleHisto}Spectral type distribution of the
    sample targets.}
\end{figure}

We constructed a joint sample of objects within 20\,pc from the two
surveys.  Both surveys together cover more than 50\% of the sky and
can be considered as almost complete in the spectral range M7--M9.5.
The full ``20pc-2MASS-DENIS'' survey contains 63 objects, 4 of which
are found in both the 2MASS and the DENIS surveys. The constructed
survey probably misses a few M7 and a few very late-M dwarfs, and at
least three of the targets are likely to be located farther away than
20\,pc.  Nevertheless, the constructed sample probably provides a
representative and robust picture of late-M dwarfs in the solar
neighborhood. We show the distribution of spectral types in the sample
in Fig.\,\ref{fig:SampleHisto}. The sample targets are given with
their apparent $J$-band magnitude and spectrophotometric distance in
Table\,\ref{tab:Sample}.

\subsection{Known Binaries}

Our sample contains four binaries that we were previously aware of;
LHS~1070B (2MASS J0024442-270825B) has a companion \citep{Leinert94}
that is about 0.2\arcsec\ away at the time of observation
\citep{Seifahrt08}. LP~349-25A (2MASS J0027559+221932A) was found to
be a binary by \citet{Forveille05}, and companions to 2MASS
J1124048+380805 and 2MASS J2206227-204706 were discovered by
\citet{Close03}. In all four systems, the spectra are dominated by the
brighter component so that we carried out our analysis in the same way
as for single stars.

During our analysis, we discovered that one of our targets, LP~775-31
(2MASS J0435151-160657), shows a double-peaked cross-correlation
profile indicative of a double-lined spectroscopic binary (SB2), i.e.,
it is a binary consisting of two nearly equally bright components.

\subsection{Observations}

Observations for our 63 sample targets were collected using the HIRES
spectrograph at Keck observatory for targets in the northern
hemisphere, and using UVES at the Very Large Telescope at Paranal
observatory for targets in the southern hemisphere (PIDs 080.D-0140
and 081.D-0190).  HIRES data were taken with a 1.15\arcsec\ slit
providing a resolving power of $R = 31,000$. The three HIRES CCDs
cover the spectral range from 570 to 1000\,nm in one exposure.  UVES
observations were carried out using a setup centered at 830\,nm
covering the spectral range 640--1020\,nm.  All data provide the
H$\alpha$ line as well as the molecular absorption bands of molecular
FeH around 1\,$\mu$m that are particularly useful for the
determination of rotation and magnetic fields in M dwarfs. These will
be investigated in a second paper.

Data reduction followed standard procedures including bias
subtraction, 2D flat-fielding, and wavelength calibration using ThAr
frames. HIRES data was reduced using the MIDAS echelle environment.
For the UVES frames, we used the ESO CPL pipeline, version 4.3.0.
There are two important differences to the standard pipeline products.
First, the version of the pipeline used in Period 81 produced
wavelengths that are offset from the correct solution by a constant
factor of 1\,\AA, this bug has been fixed in version 4.3.0. Second,
the standard pipeline extracts the spectra before flat-fielding, which
is not appropriate for spectra with strong fringing patterns. The new
pipeline version allows to flat-field the original 2D frames so that
fringe patterns are effectively removed.

The apparent brightness of our targets are in the range $J =
8.9$--$13.4$\,mag. Exposure times between 200\,s and 3\,h yielded
signal-to-noise ratios between 20 and 100 at 1\,$\mu$m in all objects.

\section{Measurements}

\subsection{Li Absorption}

\begin{figure}
  \plotone{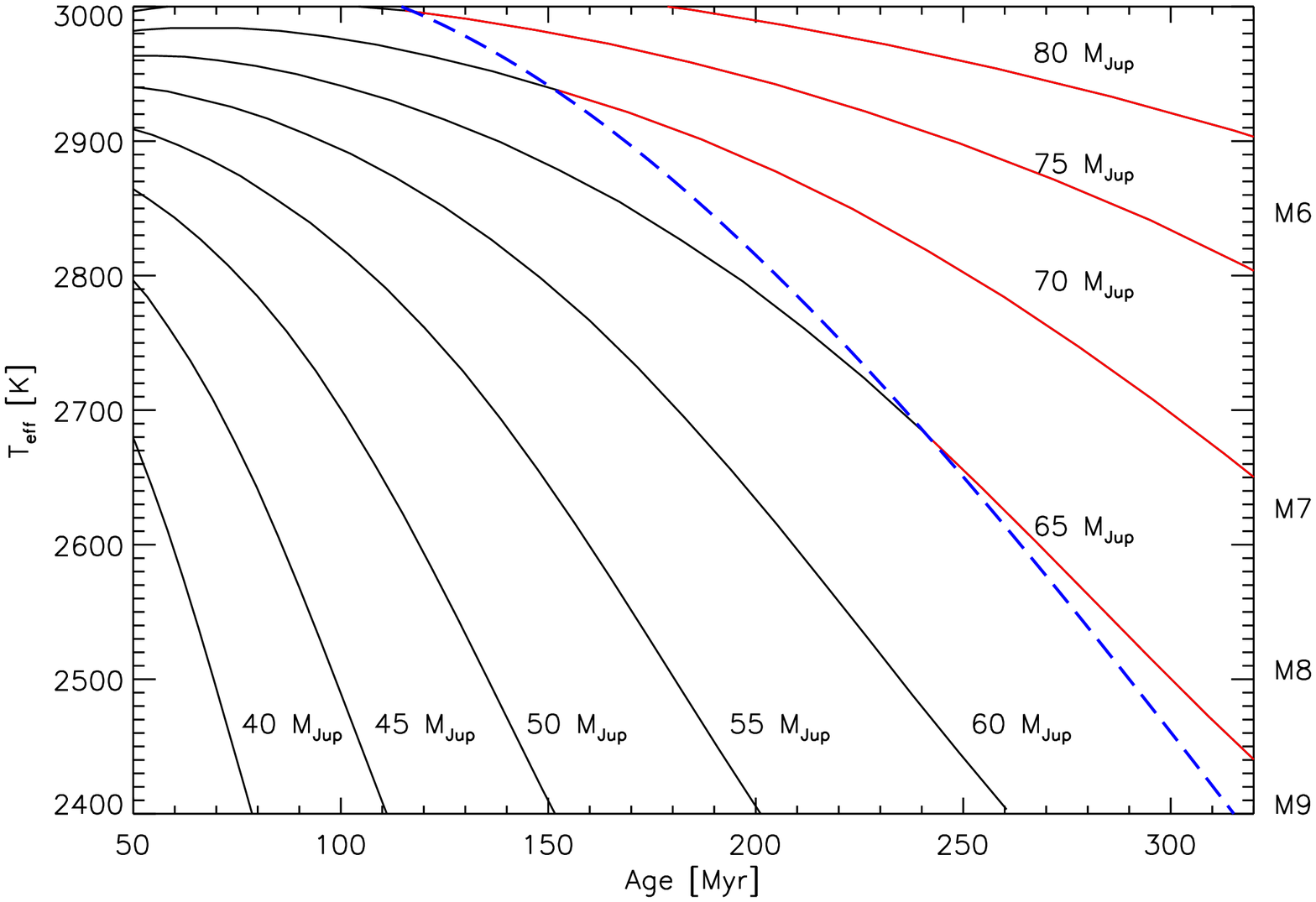}
  \caption{\label{fig:Li_Dantona}Evolutionary tracks of low mass stars
    and brown dwarfs from \citet{DAntona97}, we chose the 1998 updated
    models with a Deuterium abundance $X_D = 10^{-5}$
    \citep{Censori98}. Tracks are shown in red where the Li abundance
    is depleted by 99\% and more.  The dashed blue line divides the
    plot into regions where Li is not yet depleted (left of the dashed
    line) and it is depleted (right). Approximate spectral types
    according to \citet{Golimowski04} are indicated on the right axis.
    Note that this plot covers the mid- and late-M dwarfs and that the
    L dwarf regime is not shown.}
\end{figure}

Our spectra cover the 6708\,\AA\ Li line that can be used for the
``lithium-test'' in low-mass objects \citep[e.g.,][]{Magazzu91,
  Basri00}. In our spectra, the detection limit of the Li line is
around 0.5\,\AA. While stars quickly deplete Li on a timescale of
100\,Myr and less, brown dwarfs need longer for this process or do not
entirely deplete Li at all. We show evolutionary tracks from
\citet{DAntona97} in Fig.\,\ref{fig:Li_Dantona}. The dashed blue line
marks the region where Li is depleted by 99\,\%, i.e., objects with
detected Li lie on the left hand side of this line. For the
temperatures of our sample stars, $2400$\,K$ \la T_{\rm eff} \la
2800$\,K, this means that objects with Li are less massive than about
65\,M$_{\rm J}$, and younger than $\sim 0.5$\,Gyr. This also means
that they are less massive than the substellar limit at 75\,M$_{\rm
  J}$, hence they are young brown dwarfs.

\begin{figure}
  \parbox{\textwidth}{
    \mbox{
      \mbox{\includegraphics[width=.3\hsize]{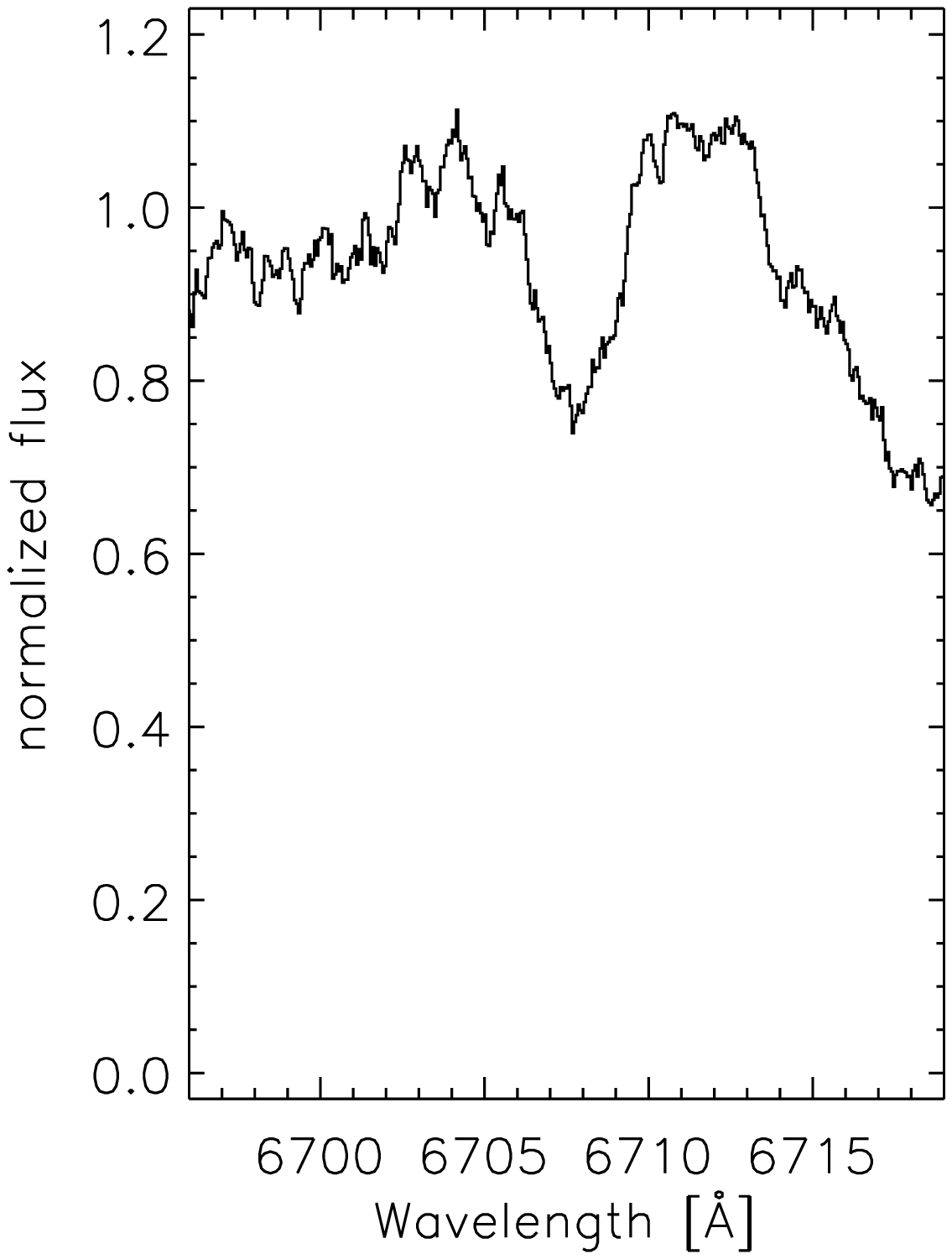}}
      \mbox{\includegraphics[width=.3\hsize]{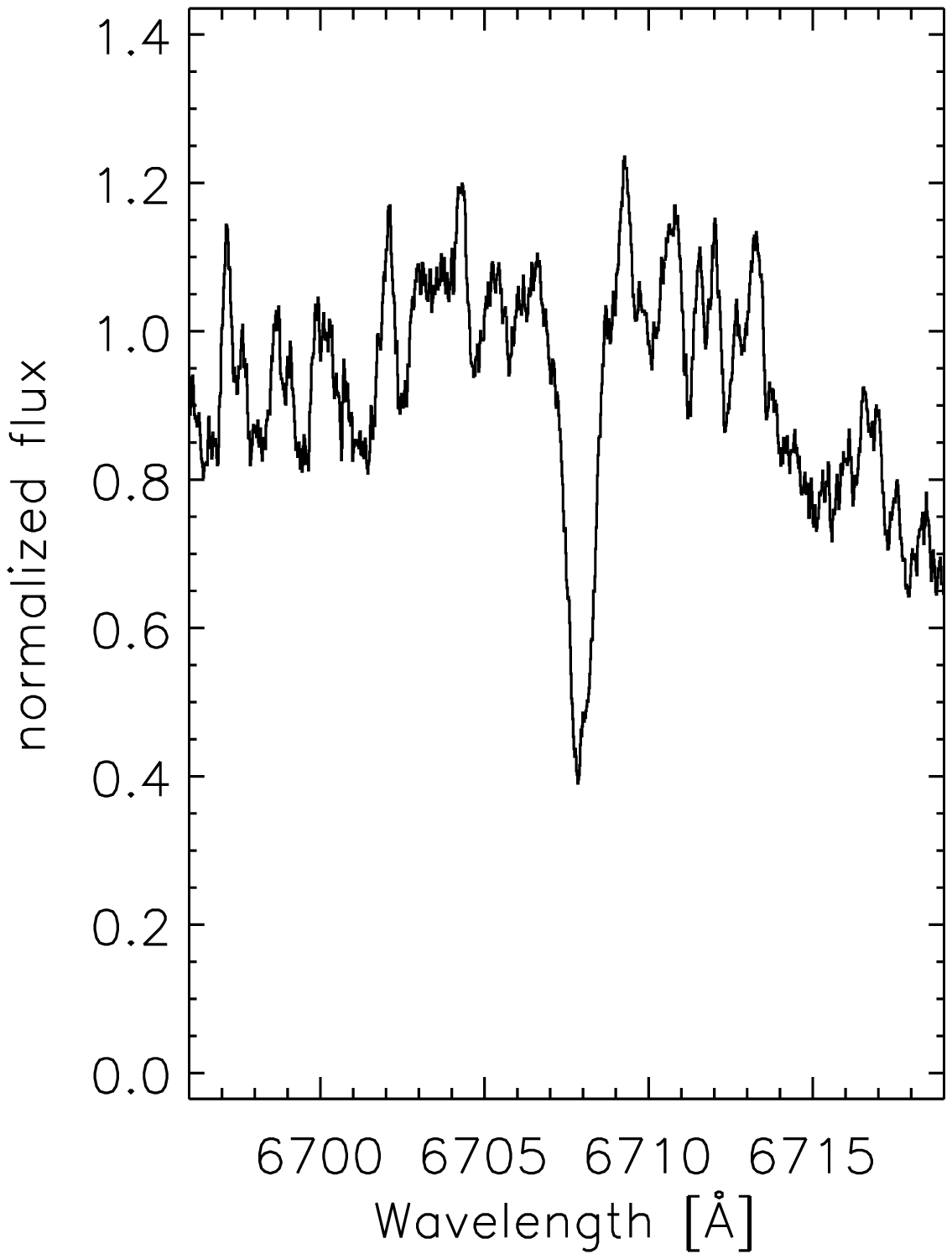}}
      \mbox{\includegraphics[width=.3\hsize]{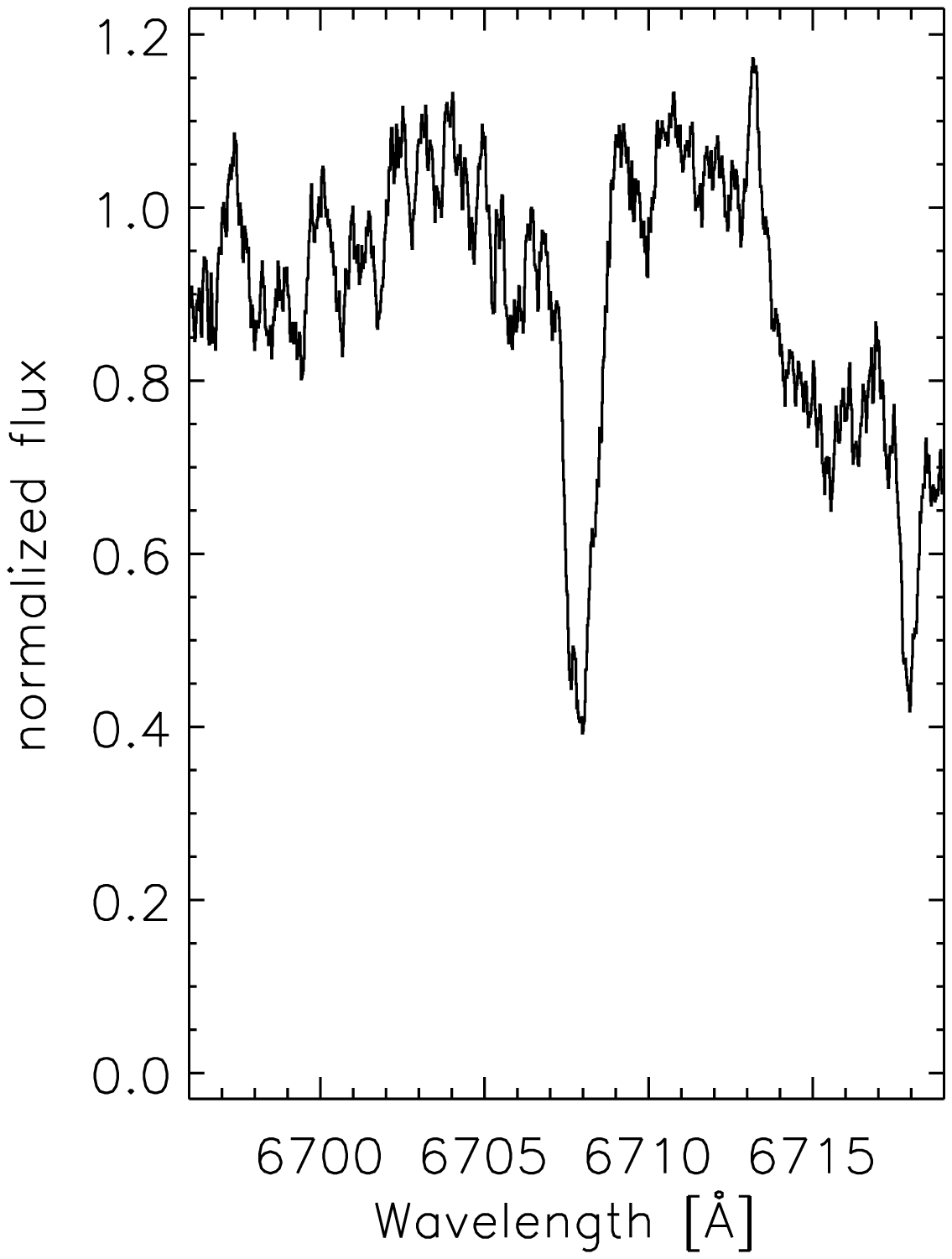}}
    }\\[1ex]
    \mbox{
      \mbox{\includegraphics[width=.3\hsize]{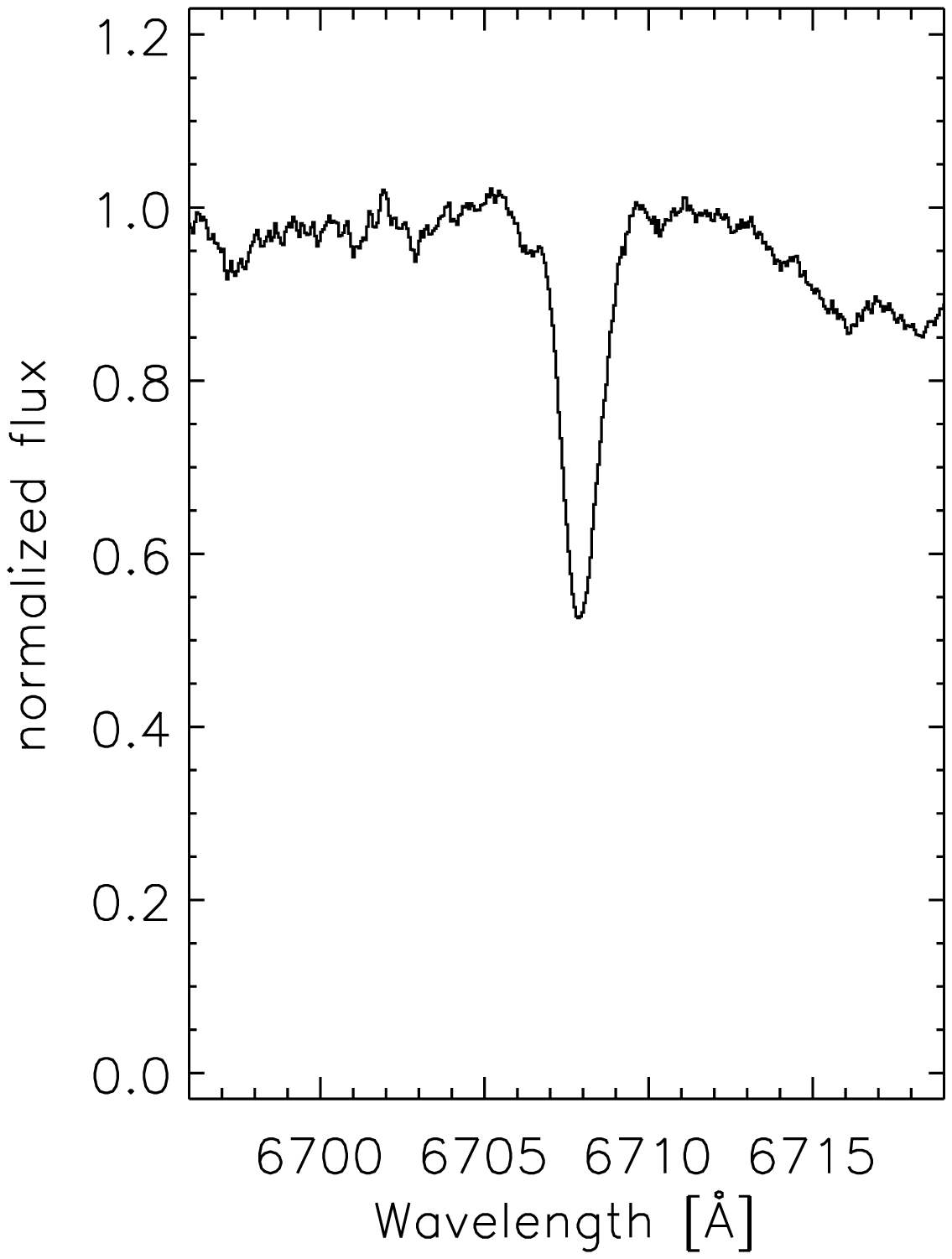}}
      \mbox{\includegraphics[width=.3\hsize]{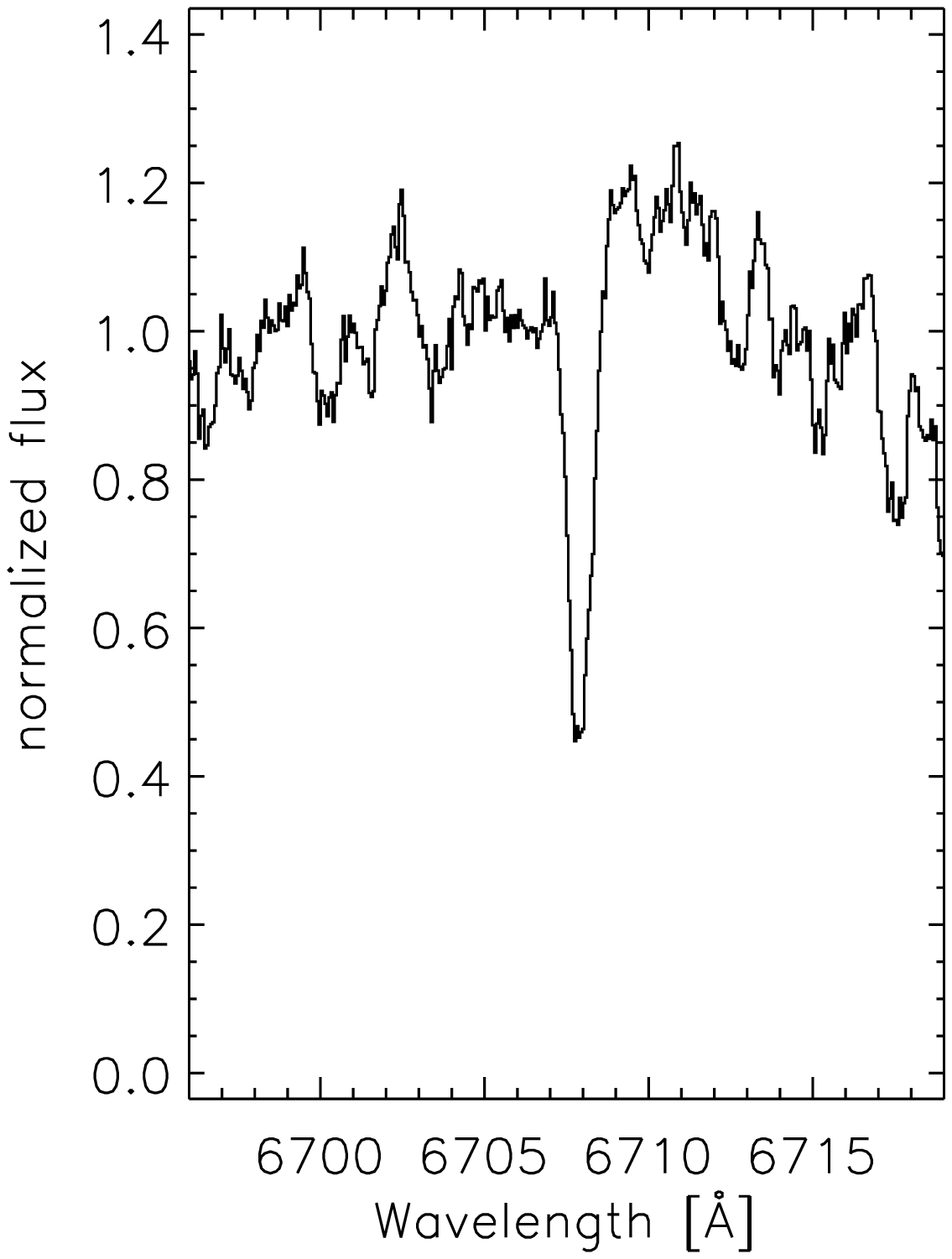}}
      \mbox{\includegraphics[width=.3\hsize]{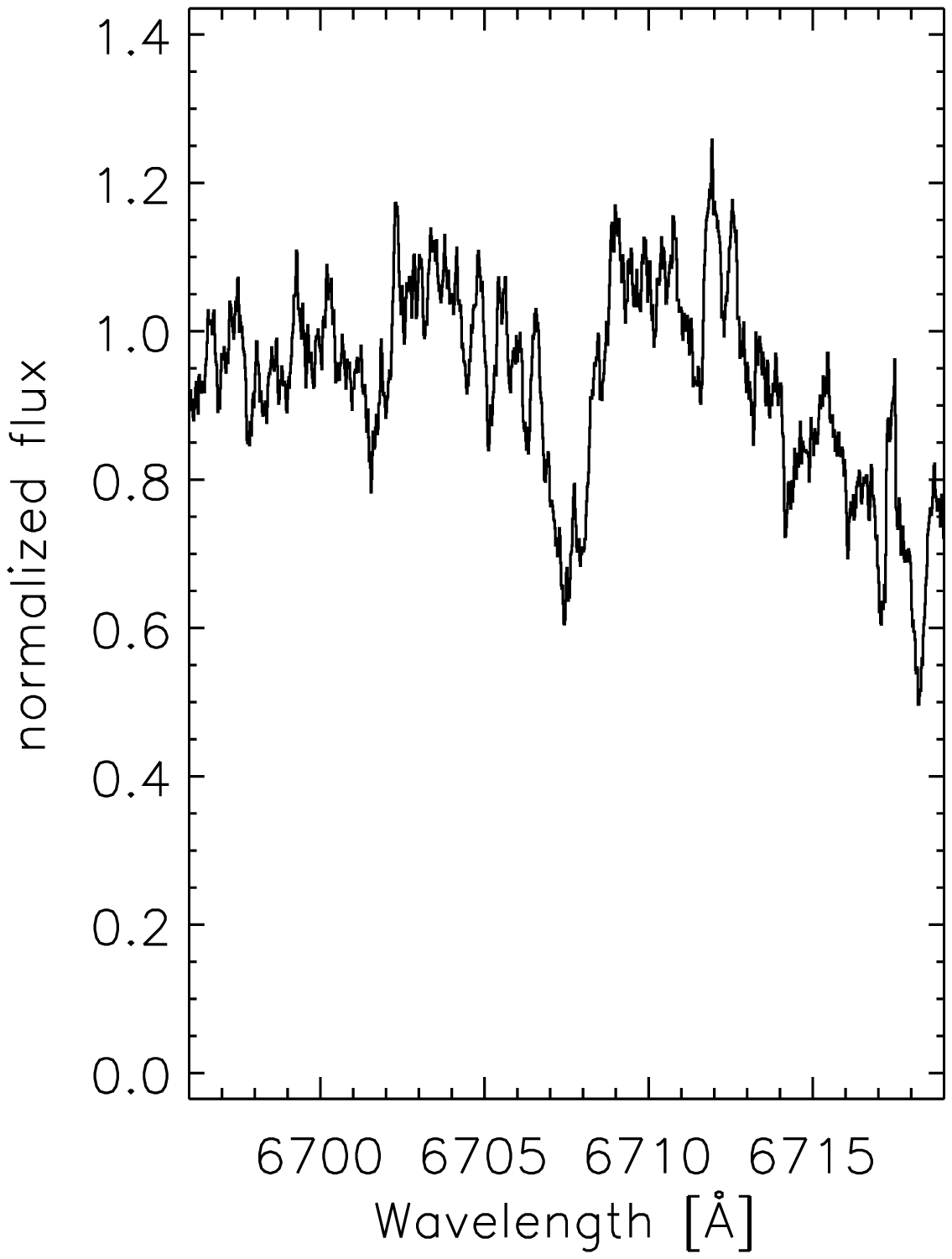}}
    }
  }
  \caption{\label{fig:Li}Li lines, from left to right: \emph{top row:}
    2MASS~J0019262$+$461407 (M8.0, note that rapid rotation causes the
    broad and relatively shallow Li line), 2MASS~J0041353$-$562112
    (M7.5)), 2MASS~J0123112$-$692138 (M8.0); \emph{bottom row:}
    2MASS~J0339352$-$352544 (M9.0), 2MASS~J0443376$+$000205 (M9.0),
    2MASS~J1411213$-$211950 (M9.0). Spectra have been smoothed with a
    15-pixel box-car for better visibility.}
\end{figure}

We have detected Li in six of the 63 objects in our sample, i.e.,
about 10\,\% of our sample are brown dwarfs younger than about
0.5\,Gyr. The Li lines of all six objects are shown in
Fig.\,\ref{fig:Li}. Only one of the targets, LP\,944$-$20
(2MASS~J0339352$-$352544), was previously known as a young brown dwarf
through the detection of its Li line \citep{Tinney98}. Youth and brown
dwarf nature of the other five objects were unknown before.
\citet{Cruz07} classified 2MASS~J0443376$+$000205 as a low gravity,
hence probably young, object.  This classification is supported by our
Li detection.

\subsection{Radial Velocities and Space Motion}
\label{sec:UVW}

Radial velocitites were measured relative to a spectrum of a standard
star. We used Gl~406 as radial velocity standard \cite[$v_{\rm rad} =
19\pm1$km\,s$^{-1}$,][]{Martin97} using the cross-correlation
technique and correcting for barycentric motion. We corrected for
wavelength calibration inconsistencies by cross-correlating telluric
lines in the O$_2$ A-band. Thus, uncertainties in our radial
velocities are on the order of 1--3\,km\,s$^{-1}$ depending on
rotational line broadening. The uncertainties of the
spectrophotometric distance are typically on the order of $\sim
10$\,\%. For our sample of nearby objects this introduces an
uncertainty in the space motion components of typically $\sim
1$\,km\,s$^{-1}$. If an object is substantially younger than assumed
for the spectrophotometric distance calculation, the error is much
larger. This effect is illustrated for 2MASSJ~0041353-562112 in
\cite{RE09}, but is probably not important in the other objects.

Space motions $U, V$, and $W$ were computed using the IDL procedure
{\ttfamily gal\_uvw}\footnote{\ttfamily
  http://idlastro.gsfc.nasa.gov/contents.html}, but we use a
right-handed coordinate system with $U$ towards the Galactic center.
For most of our targets, proper motions are reported in
\citet{Faherty09}. \citet{PhanBao01} and \citet{PhanBao03} measured
proper motions for some of our objects, and we use their values where
available. One object, LHS~1070B (2MASS J0024442-270825B), is not
included in either of the works, and we use the proper motion given in
\citet{Salim03}.  All radial velocities and space motion vectors [$U,
V, W$] together with the references for distance and proper motion are
given in Table\,\ref{tab:Sample}.

\section{Results}

\subsection{Lithium Brown Dwarfs}

From our volume-limited sample, we can draw robust constraints on the
fraction of young brown dwarfs among the ultracool dwarfs. Our results
are summarized in Table\,\ref{tab:Li}. \citet{Reid02a} found fractions
of $F_{\rm{Li}} = (6 \pm 4)\%)$ for M7--M9.5 dwarfs and $F_{\rm{Li}} =
(10 \pm 7)\%$ for M8--M9.5 dwarfs.  The Li fractions from our sample
are somewhat higher than the ones reported in \cite{Reid02a}, but the
discrepancy does not exceed the uncertainties due to the small sample
size.

We conclude that the fraction of lithium brown dwarfs among M7--M9.5
dwarfs is at most only little higher than previously reported. This
means that the mass function index $\alpha$ could perhaps be a little
higher than reported in \citet{Reid02a}. Nevertheless, their
conclusion that $\alpha < 2$ for both the Arizona and Lyon models is
still valid.

\citet{Kirkpatrick08} investigated the fraction of objects with Li
detections among L dwarfs together with a search for objects with
spectroscopic signatures of youth, i.e., low gravity.  They show that
the fraction of Li detections is about 10\,\% at spectral type L0 and
rises to about 50--60\,\% at mid-L (see their Fig.\,17).  After L6,
the fraction turns over and declines towards the late-L types probably
because Li disappears into molecules \citep{Lodders99}.
\citet{Kirkpatrick08} emphasize that Li is probably \emph{not}
detectable in \emph{very} young L dwarfs, i.e., during the first few
ten million years, because the Li line is not visible at very low
gravity.

\begin{figure}
  \plotone{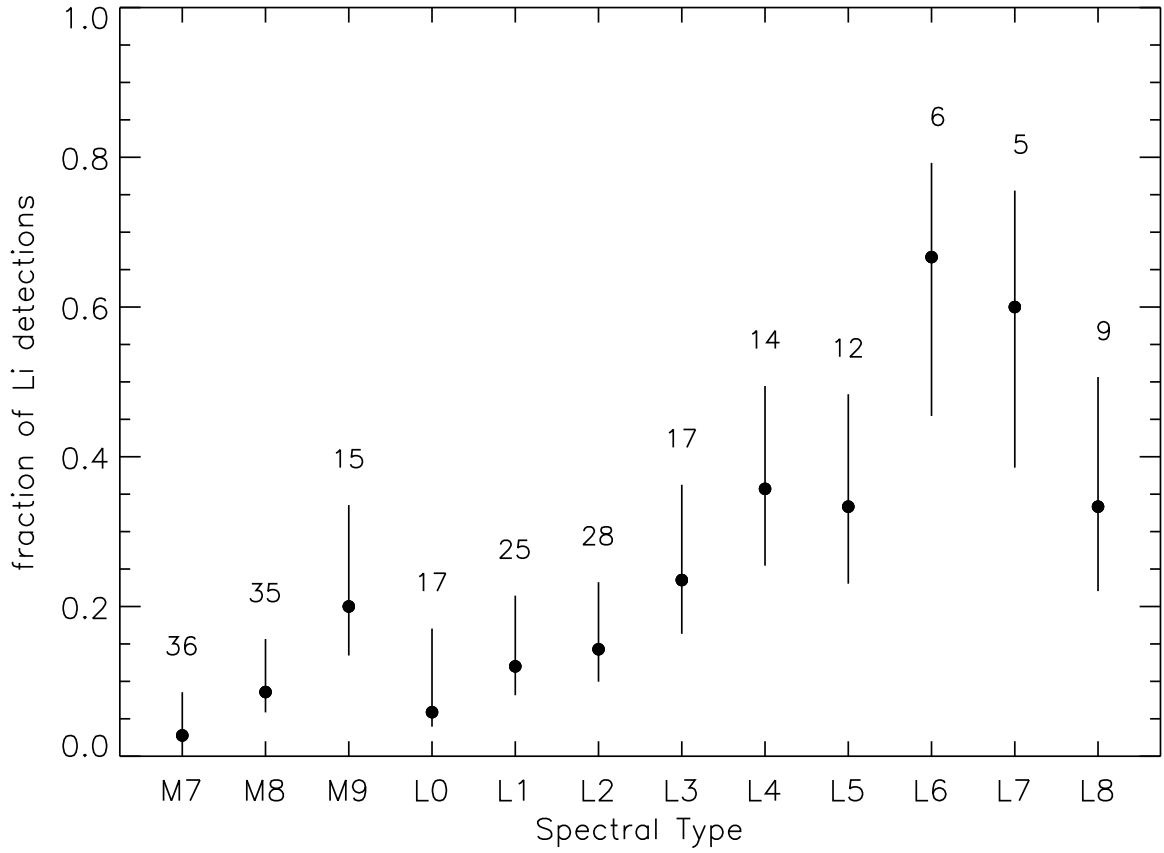}
  \caption{\label{fig:BDfrac}Fraction of objects with detected Li per
    spectral type. M7--M9 objects are from this work and
    \citet{Reid02a}, L dwarfs are from \citet{Reiners08} and
    \citet{Kirkpatrick99, Kirkpatrick00, Kirkpatrick08}. Numbers
    denote the total number of objects per spectral bin, error bars
    are 1$\sigma$ uncertainties.}
\end{figure}

We plot the fraction of lithium brown dwarfs as a function of spectral
types M7--L8 in Fig.\,\ref{fig:BDfrac}. M7--M9 dwarfs are from this
work and from \citet{Reid02a}, L dwarfs are taken from
\citet{Reiners08}, and from the Li analysis of \citet{Kirkpatrick08}
including data from \citet{Kirkpatrick99, Kirkpatrick00}. While the
Kirkpatrick et al. works employ low-resolution LRIS spectra, the other
Li detections are based on spectra with higher spectral resolution.
Nevertheless, the detection thresholds from the different samples are
not very different because in L dwarfs it is usually not the spectral
resolution that limits the detection of Li (note that the Li lines
shown in Fig.\,\ref{fig:Li} are smoothed to a lower effective
resolution). Comparing the subsamples taken from high- and
low-resolution spectra, we see no difference in the fraction of Li
detections, and the results for objects that are contained in both
samples are consistent.

The fraction of lithium brown dwarfs among ultracool dwarfs
continually rises from M7 to M9, because late-M dwarfs in general are
less massive, so that the fraction of objects with $M \la
0.07$\,M$_{\odot}$ is growing towards later spectral types. In
Fig.\,\ref{fig:BDfrac}, there is also a general trend that the Li
fraction rises from M7 to mid-L, which is consistent with a higher
brown dwarf fraction at later spectral type. Note that the time
required to deplete Li in high-mass brown dwarfs ($\ga 60$\,M$_{\rm
  J}$) is longer among the L dwarfs ($\sim 1$\,Gyr) than at late-M
spectral type ($\sim300$\,Myr). However, the fraction of Li detections
among M9 dwarfs is much higher than at M8, and, in particular, it is
much higher than at spectral types L0 and L1.  The number of objects
in this sample is still fairly low, and the results are formally still
consistent with a smooth rise in the range M7--L6 if the M9 bin is
interpreted as an outlier. On the other hand, the discontinuity at
spectral type M9 is fairly high while the overall scatter between the
other bins is rather small. Three out of the 15 objects at spectral
type M9 and M9.5 show Li, all three are most likely within 20\,pc. The
statistical significance of the M9 dicontinuity is comparable to that
of the apparent downturn in Li at late L types noted by \citet{
  Kirkpatrick08}.  In the following, we discuss possible explanations.

The first possible explanation is that our combined sample is not
strictly volume-limited, in particular the L dwarfs are not selected
according to their distance. This implies a bias towards young objects
and probably causes an excess of young dwarfs that still have Li.
Most distances of our M dwarf sample are taken from spectrophotometry,
which means that a few other young objects contained in our M dwarf
sample may be situated outside 20\,pc. However, this problem should
apply even more to the early-L dwarfs, in which no effort was made to
include faint old objects out to a certain distance. Thus, one would
expect the early-L dwarfs to be biased towards higher fraction of Li
detections, and this bias should be higher than the bias among the
late-M dwarfs. Furthermore, the fraction of Li detections in our
M9/M9.5 targets from the volume-limited sample is higher than the
fraction in the full sample including the targets from
\citet{Reid02a}. Thus, an observational effect that causes a bias
towards young objects particularly at the M9 spectral bin is unlikely
an explanation for the observed discontinuity.

A second potential reason for the discontinuity is that the L dwarf
spectra are of lower quality compared to the M dwarf spectra taken for
this work.  This would lead to a higher Li detection threshold and
hence a lower detection rate.  While such an explanation might hold
for the low-resolution spectra \citep[a detection threshold of 3\,\AA\
is reported in ][]{Kirkpatrick99, Kirkpatrick00, Kirkpatrick08}, it
does not apply to the high resolution spectra of \citet{Reiners08}. In
the latter sample alone, the ratio of Li detections to total number of
targets is 0/10 (0\,\%) at spectral type L0 and 1/14 (7\,\%) at
spectral type L1, which is an even lower ratio than the one from low
resolution spectra in \citet{Kirkpatrick08}. Thus, differences in data
quality do probably not lead to systematic differences in the Li
fraction, and it seems unlikely that the detection threshold is the
main reason for the low fraction of Li detections at early L dwarfs.

As a third potential reason, we speculate that the discontinuity is
real, i.e., the fraction of objects with detectable Li peaks at
spectral type M9. We see two possible ways to explain this. 1) Lithium
brown dwarfs with temperatures of $\sim$2200\,K may be classified as
M9, but at older ages stars of the same temperature could be
classified as L0.  Classification of spectral types between M9 and L0
is based on molecular species that are temperature dependent
\emph{and} sensitive to the presence of dust, which is strongly
influenced by gravity. In other words, the relation between
temperature and spectral class may be a function of age. 2) The
history of the star formation rate together with the detectability of
Li as a function of age \citep[see][]{Kirkpatrick08} may conspire so
that today we observe an excess of Li detections at spectral type M9.
This may imply a burst of star formation at a particular age. So far,
however, we are not convinced that either of the two is a very likely
scenario, and we will have to wait for a larger sample and better
distance estimates to see whether the maximum disappears with better
statistics.

\subsection{Space Motion of Ultracool Dwarfs}

Space motions of M dwarfs and ultracool dwarfs have raised a lot of
interest during the last decades because the velocity dispersion of a
sample can be tied to its age, which allows the study of the star
formation rate and the time-dependence of physical processes like for
example activity. \citet{Hawley96} have shown in a volume-limited
sample of M dwarfs that the velocity dispersion of active dMe stars is
smaller than the dispersion of older non-active stars. \citet{Reid02a}
determined the velocity dispersion of ultracool objects finding that
the velocity dispersion of their M7--M9.5 sample is smaller than the
dispersion in the full dM sample of \cite{Hawley96}. \citet{Reid02b}
revisit the space velocities of ultracool M dwarfs suggesting that the
early-M dwarf sample consists of members of both the thick disk and
the thin disk. Late-M dwarfs, on the other hand, may have lost their
thick disk component because the lower metallicity of these stars
moves the hydrogen burning limit to earlier spectral types so that
these stars have cooled towards later spectral types and don't show up
as M7--M9.5 objects any longer.

\begin{figure}
  \mbox{
      \mbox{\includegraphics[width=.45\hsize]{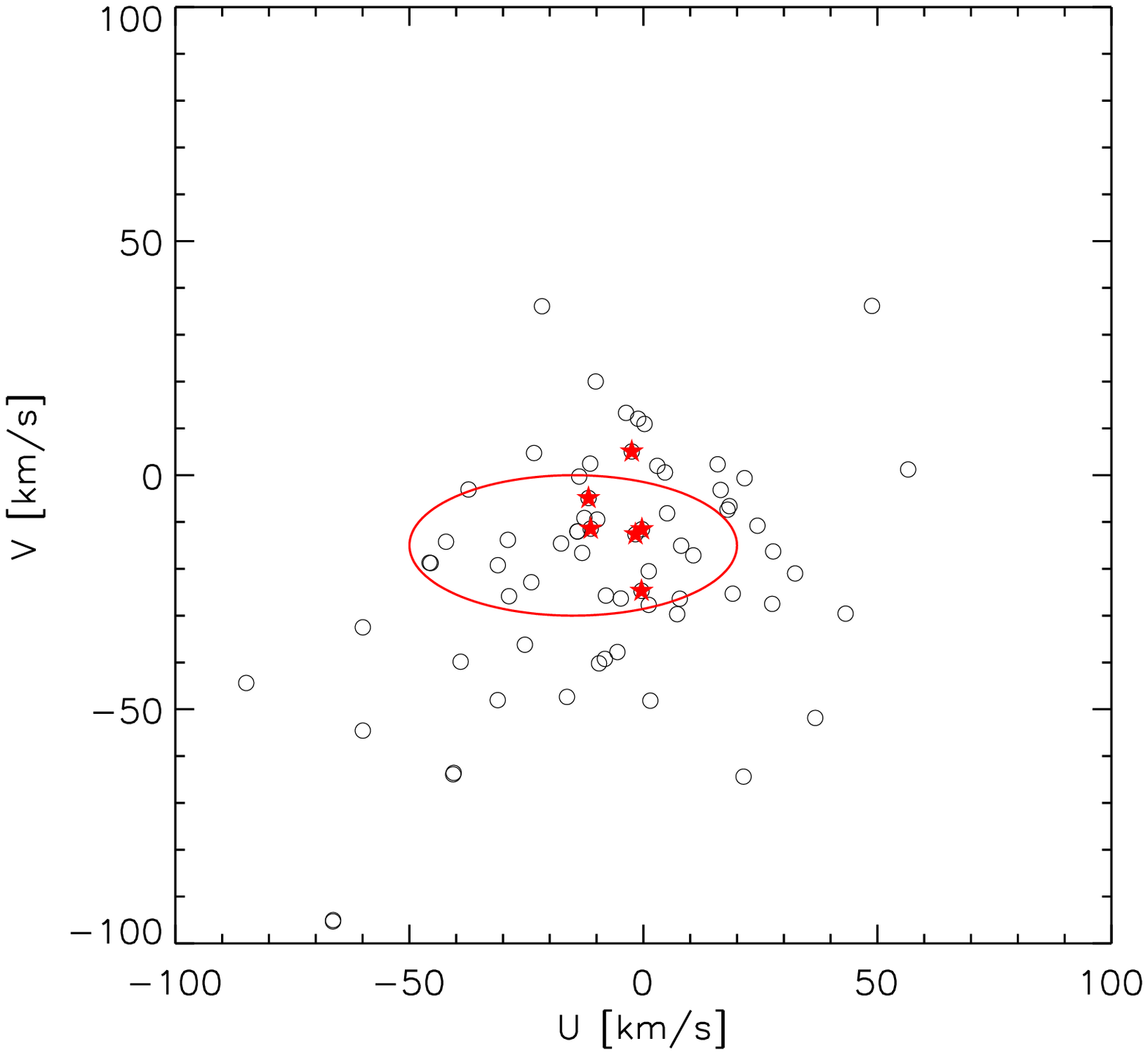}}
      \hspace{2ex}
      \mbox{\includegraphics[width=.45\hsize]{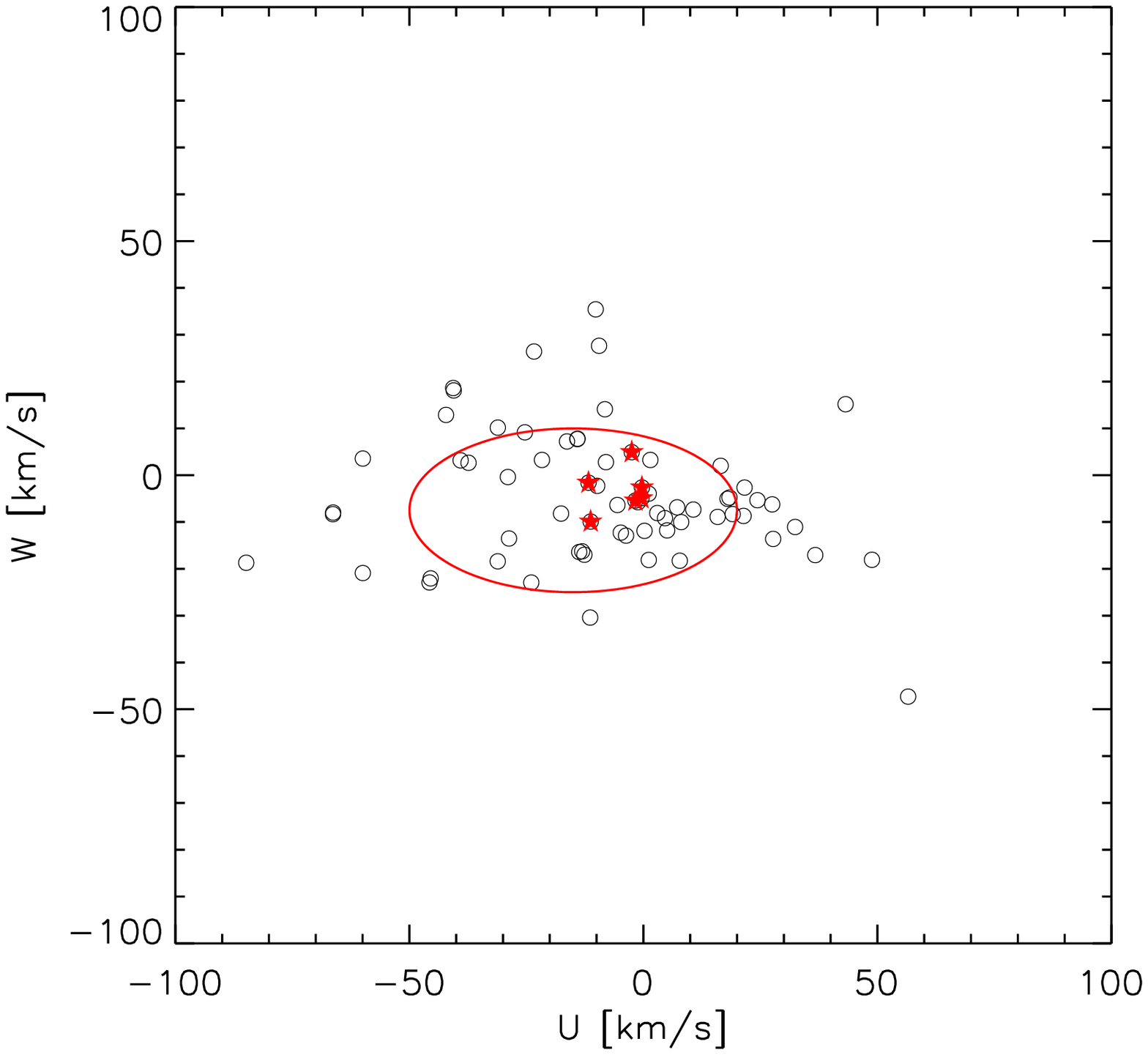}}
    }
    \caption{\label{fig:UV}Space velocities in $U-V$ (left) and $U-W$
      (right) diagrams. The thin disk is indicated as a red ellipsoid,
      and the six lithium brown dwarfs are shown as red stars.}
\end{figure}

\begin{figure}
  \mbox{
    \mbox{\includegraphics[width=.3\hsize]{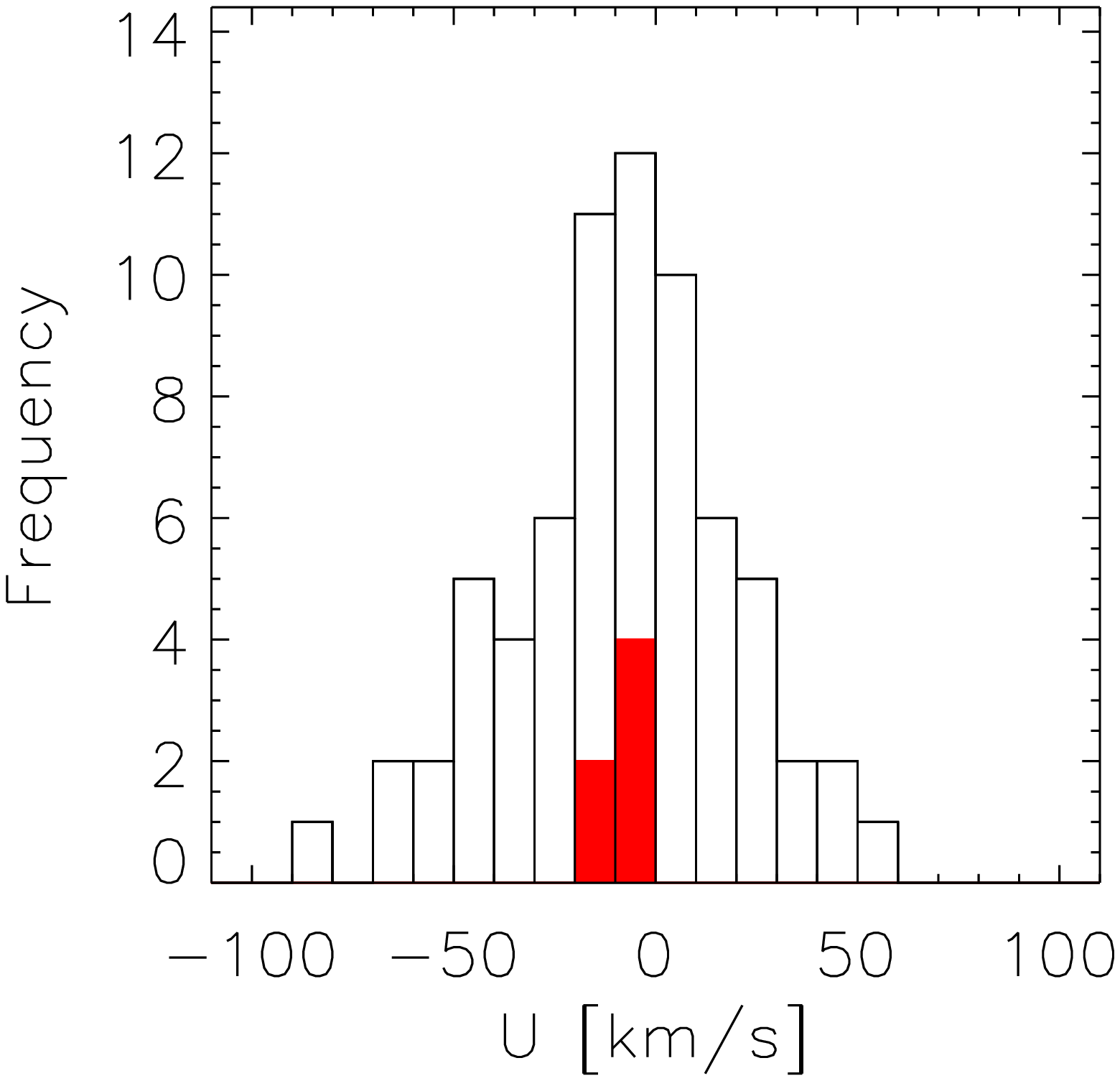}}
    \mbox{\includegraphics[width=.3\hsize]{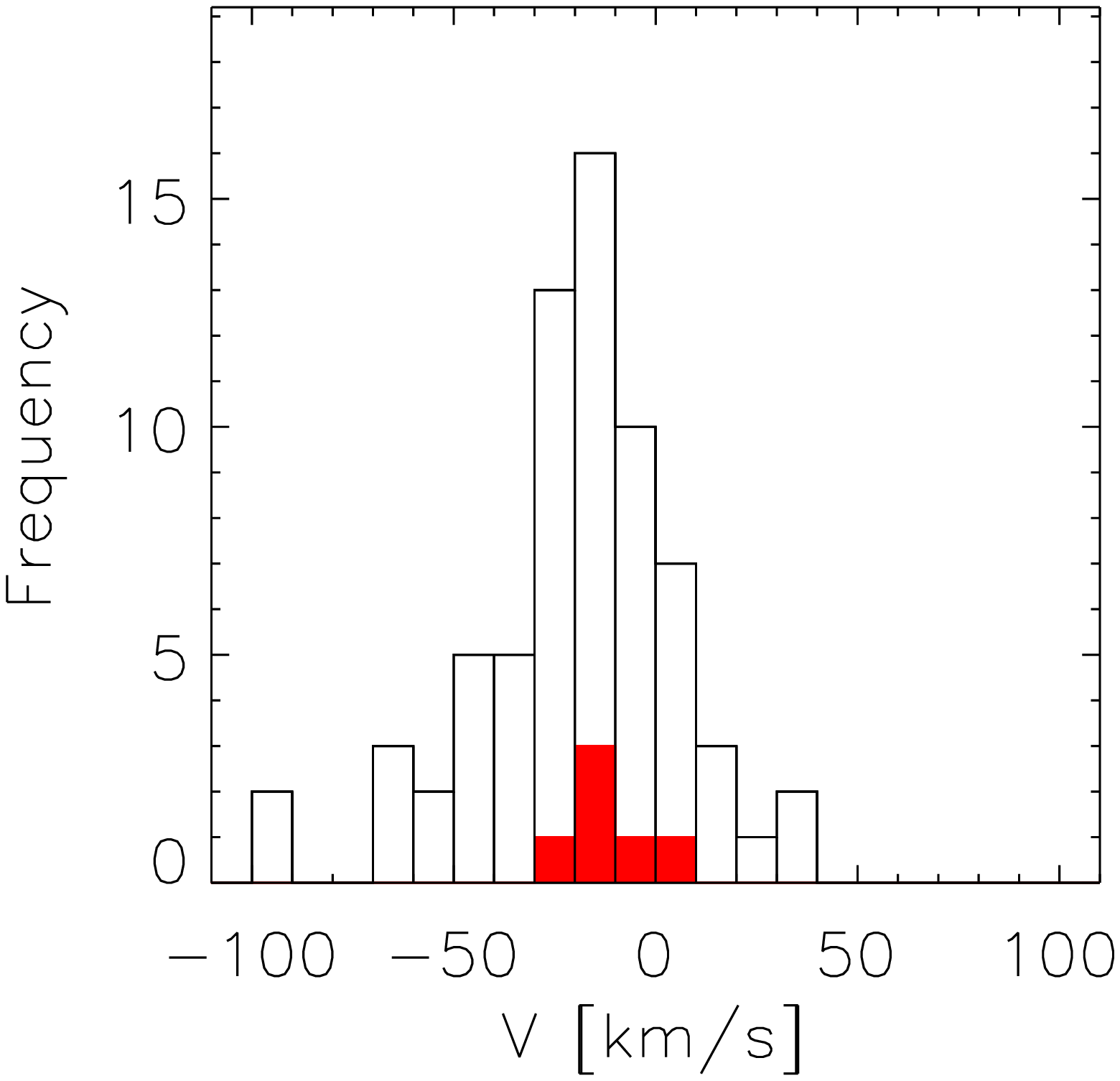}}
    \mbox{\includegraphics[width=.3\hsize]{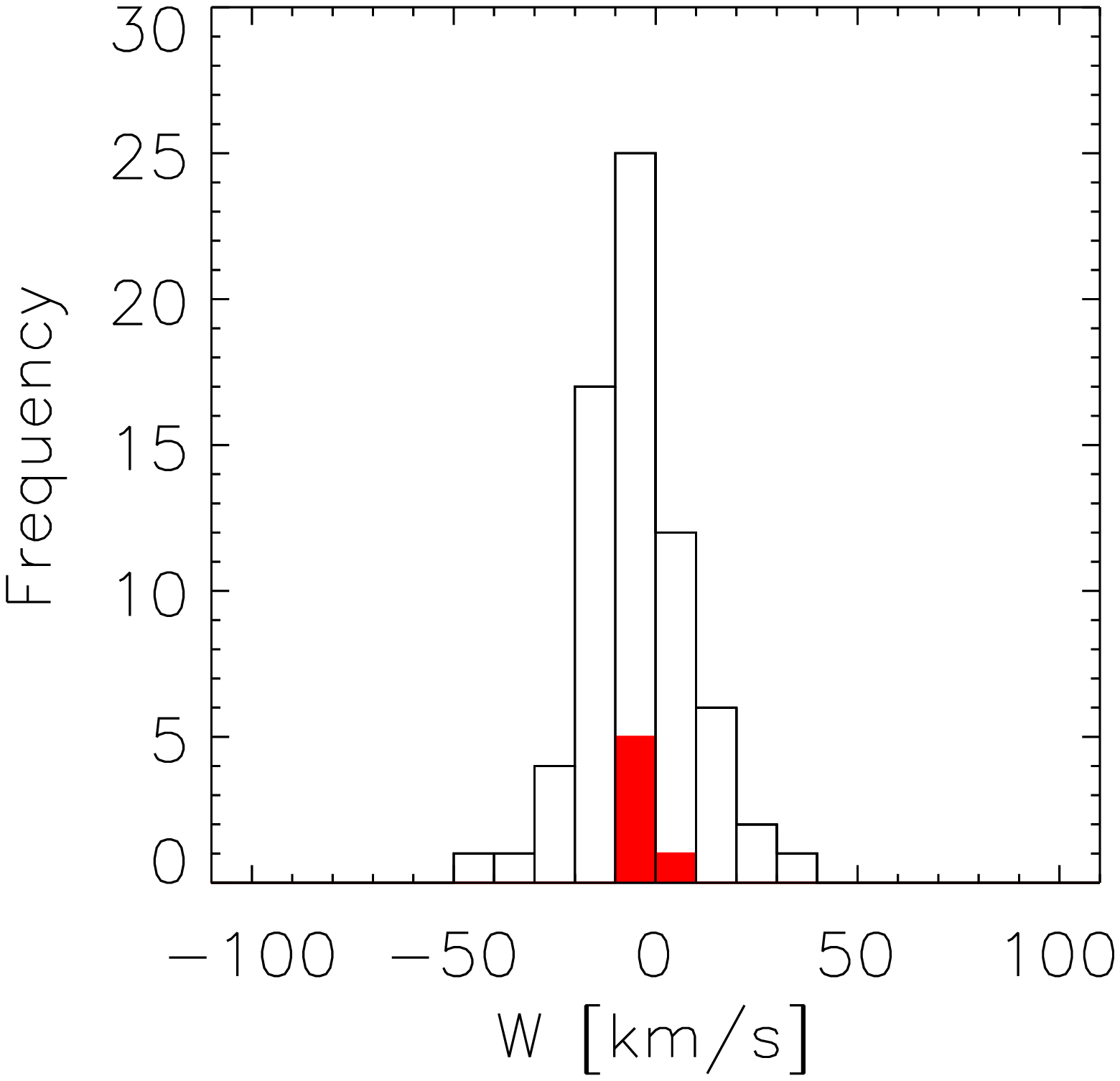}}
  }
  \caption{\label{fig:Vhisto}From left to right: Histograms of space
    motion components $U, V$, and $W$. The distribution of lithium
    brown dwarfs is overplotted as a filled red histogram.}
\end{figure}

We show the space velocities in $U-V$ and $U-W$ diagrams in
Fig.\,\ref{fig:UV}, and the distribution of $U, V$, and $W$ in
Fig.\,\ref{fig:Vhisto}. Lithium brown dwarfs are shown as red stars in
Fig.\,\ref{fig:UV} and are overplotted as filled red histograms in
Fig.\,\ref{fig:Vhisto}. The distribution of space velocities in
ultracool dwarfs is clearly concentrated around the thin disk, and the
six lithium brown dwarfs cluster in a very narrow range in the center
of the thin disk in all three space velocities. This clustering
implies a very small velocity dispersion, which is an indicator for a
young sample \citep[see][and next Section]{Wielen77}. Thus, space
velocities of lithium brown dwarfs are consistent with very low age.
In the following, we will derive the age of our sample of M7--M9.5
dwarfs.

\subsection{Ages from Velocity Dispersions}

\citet{Wielen77} has shown how a velocity dispersion of a sample of
stars can be used to determine the age of the sample. More recently,
\citet{Fuchs01} refined this picture using more recent data from
different works.

First, it is important to realize that the formulae provided by
\citet{Wielen77} are valid for dispersions $\sigma_{U}$, $\sigma_{V}$,
and $\sigma_{W}$ that are weighted by their vertical velocity $|W|$
\citep[Eqs. 1--3 in][see also Reid et al., 2002b]{Wielen77}. Although
it might be arguable whether this weighting is useful or not when
comparing dispersions from different samples, it has to be applied if
the age of a sample should be determined from the formulae in
\citet{Wielen77}. The dispersion of space velocities in our sample is
$(\sigma_U, \sigma_V, \sigma_W) = (27.2, 22.6, 14.6)$\,km\,s$^{-1}$,
and the $|W|$-corrected velocity dispersions are $(\sigma_U, \sigma_V,
\sigma_W) = (30.7, 22.6, 16.0)$\,km\,s$^{-1}$.

A second but much more crucial principle for the calculation of ages
from velocity dispersions is the calculation of the total velocity
dispersion,
\begin{equation} 
  \sigma_{\rm tot} = (\sigma_U^2 + \sigma_V^2 + \sigma_W^2)^{1/2}.
\end{equation}
This means that $\sigma_{\rm tot}$ is \emph{not} the dispersion of the
total velocity scalar $v_{\rm tot} = (U^2 + V^2 + W^2)^{1/2}$, since
the latter will always be smaller because motions
into different directions partially cancel out each other (for
example, a sample with velocities equally distributed into all
directions but with the same total velocity will have zero dispersion
around $v_{\rm tot}$).  The uncorrected total dispersion of our sample
is $\sigma_{\rm tot} = 38.3$\,km\,s$^{-1}$, and the $|W|$-corrected
total dispersion is $\sigma_{\rm tot} = 41.3$\,km\,s$^{-1}$.

It should be noted that, following the same argumentation as for
$\sigma_{\rm tot}$, the total dispersion in the tangential velocity
scalar, $v_{\rm tan}$, cannot be applied to calculate $\sigma_{\rm
  tot}$ using $\sigma_{\rm tot} = (3/2)^{1/2} \sigma_{\rm tan}$: The
total dispersion in the two-dimensional velocity $v_{\rm tan}$ must be
the squared sum of the two individual dispersion components. Thus, the
total dispersion in $v_{\rm tan}$ cannot be used for the calculation
of age from the Wielen-relation.

Recently, kinematics of ultracool dwarfs and brown dwarfs have been
presented by \citet{Schmidt07}, \citet{Zapatero07}, and
\citet{Faherty09} using the velocity dispersion of a sample of stars
to calculate the age of the sample following the work of
\citet{Wielen77}. \citet{Schmidt07} derives an age of 3.1\,Gyrs for a
sample of late-M and L dwarfs, \citet{Zapatero07} finds ages of
4\,Gyrs for their late-M dwarfs and about 1\,Gyr for L and T dwarfs,
and \citet{Faherty09} reports ages between 2 and 3\,Gyrs for late-M,
L, and T dwarfs. Unfortunately all three of these papers contain the
errors in interpretation of the total velocity dispersion pointed out
in the previous two paragraphs.

Once the velocity dispersion $\sigma_{\rm tot}$ is determined for a
sample, it can be translated into an age according to
\citet{Wielen77}. He presents three different equations to carry out
this task, and it is often difficult to find out which prescription
authors use when publishing an age for a sample of stars.  The first
prescription assumes a constant diffusion coefficient and is
apparently not much used in recent work \citep[Eq.\,(8)
in][]{Wielen77}. The two other formulae assume velocity-dependent
diffusion coefficients. The two formulae, Eqs.\,(13) and (16) in
\citet{Wielen77}, differ in the way they treat the diffusion
coefficient $C_v$, they are
\begin{eqnarray}
  \label{Eq:2}
  \sigma_v^3 & = & \sigma_{v,0}^3 + \frac{3}{2}\gamma_v\tau \\
  \label{Eq:3}
  \sigma_v^3 & = & \sigma_{v,0}^3 + \frac{3}{2}\gamma_{v,p}T_\gamma(\exp{(\tau/T_\gamma)} - 1 ),
\end{eqnarray}
with $\tau$ the age of the sample in yr, $\sigma_{v,0} =
10$\,km\,s$^{-1}$, $\gamma_v = 1.4~10^{-5}$\,(km s$^{-1}$)$^3$/yr,
$\gamma_{v,p} = 1.1~10^{-5}$\,(km s$^{-1}$)$^3$/yr, and $T_\gamma =
5~10^9$\,yr. According to \citet{Wielen77}, Eq.\,\ref{Eq:2} is
inadequate for ages above 3\,Gyrs, which means that Eq.\,\ref{Eq:3}
should be preferred\footnote{\cite{Schmidt07} use Eq.\,(\ref{Eq:2})
  and present it in a slightly different form with $t$ the age of the
  sample and $\tau$ a characteristic timescale, $\tau = 2~10^8$\,yr.
  However, this actually should have been written as 
$\tau^{-1} \approx 2~10^{-8}$\,yr$^{-1}$. It provides a fortunate
compensation for the error in computing total velocity dispersions.
 Thus the results of \citet{Schmidt07}, \citet{Zapatero07}, and
\citet{Faherty09} end up fortuitously consistent with other work 
(and ours). To illustrate, \citet{Schmidt07} find an age of 3.1\,Gyr 
for $\sigma_{\rm  tot} = 20.8$\,km\,s$^{-1}$, but this low a velocity 
dispersion would in fact imply an age well below 1\,Gyr. }.

\begin{figure}
  \plotone{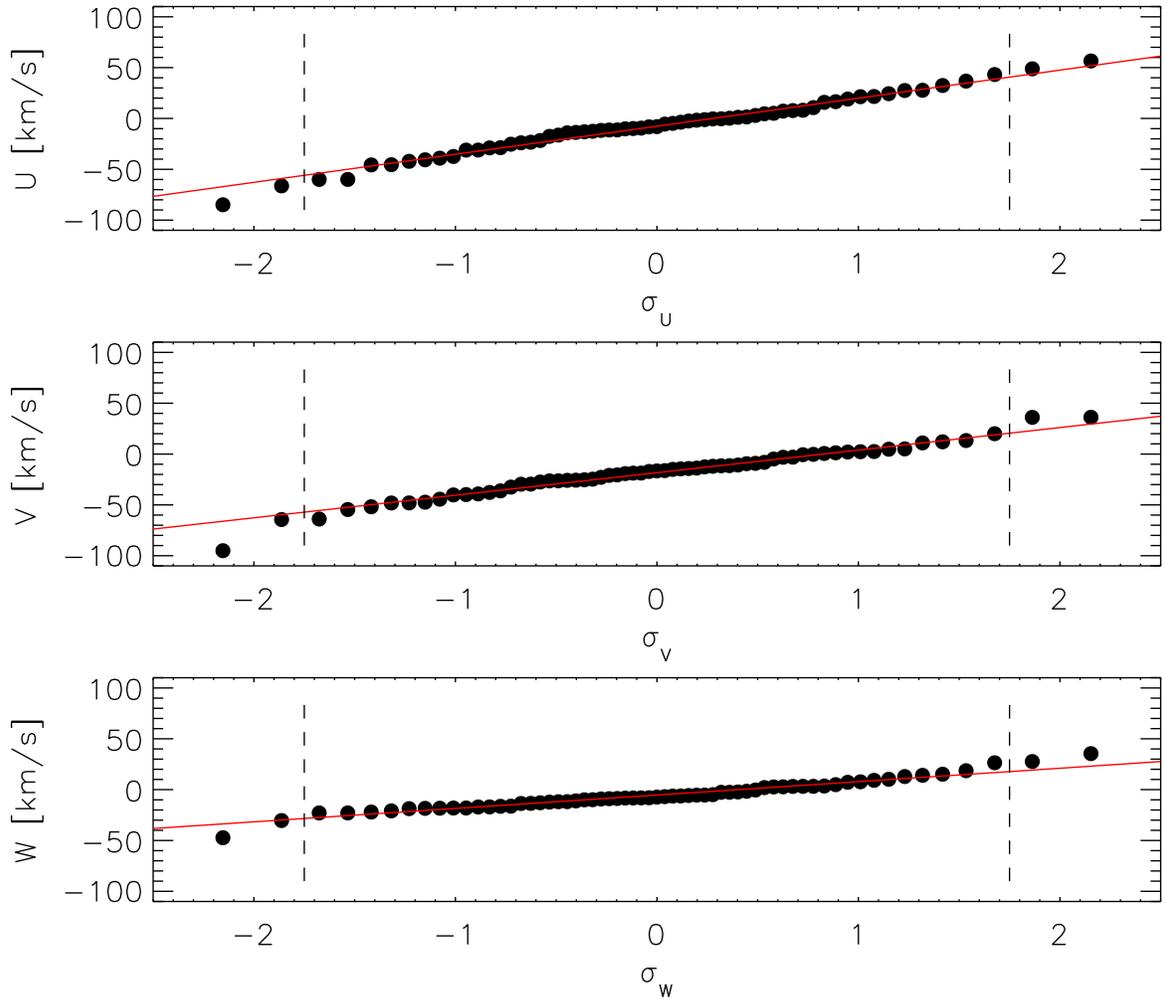}
  \caption{\label{fig:probit}Probability plots in $U, V$, and $W$. The
    red line is a fit to the data within $\pm1.75\sigma$ of the
    expected distribution assuming the data were following a Gaussian
    distribution.}
\end{figure}

The determination of an age from a stellar sample assumes that the
sample members (at least in a very qualitative sense) fulfill the
requirement of homogeneity, i.e., their velocity distribution should
not differ drastically from a Gaussian distribution. A robust way to
investigate the velocity distribution in a sample is to use
probability plots, or ``probit'' plots \citep{Lutz80}. In a
probability plot, the data points are sorted and assigned a
probability bin, which is the distance to the mean in units of the
standard deviation, that each of the sorted points in a strictly
Gaussian distribution would have. If a distribution is Gaussian, the
sample points will follow a straight line, and the value of the
standard deviation, $\sigma$, is the steepness of this line. We show
the probability plots for our sample in Fig.\,\ref{fig:probit}. The
largest part of the sample relatively well resembles a
well-constrained line, which means that except for a only a handful of
outliers, the distribution of space velocities is consistent with a
Gaussian distribution. We overplot a fit to the inner $\pm 1.75\sigma$
of the velocity distribution in Fig.\,\ref{fig:probit}. Velocity
dispersions and ages derived from the inner part of this distribution
are statistically indistinguishable from results taken from the full
sample.

From the $|W|$-corrected space velocity dispersions we have calculated
the age of our sample as described above. Our results are summarized
in Table\,\ref{tab:Ages}, the age estimate for our volume-limited
sample of ultracool dwarfs from this methodology is $\tau = 3.1$\,Gyr.
We compare this result to the sample presented in \citet{Reid02a}.
These authors also show that their sample matches a Gaussian
distribution quite well. We calculated $|W|$-corrected space velocity
dispersions from the 32 M7--M9.5 dwarfs (cp.\ their Eq.~(3)) given in
their Table\,5, and from the Wielen-relation we derive an age for the
sample of $\tau = 3.1$\,Gyr. Given the substantial uncertainties of
this age-determination, which are on the order of several hundred Myr,
this is in remarkable agreement with the age derived from our sample.
Thus, there is good evidence that the mean age of the local ultracool
dwarf population is about 3\,Gyr.

\section{Summary}

Ultracool dwarfs of spectral types M7--M9.5 are of particular interest
to our understanding of the physics of low-mass stars and brown
dwarfs. In order to investigate the physics of ultracool dwarfs, we
constructed a volume-limited sample of 63 M7--M9.5 dwarfs from the
DENIS and 2MASS samples. In this first paper, we present the sample
composition, the fraction of young brown dwarfs from detections of Li,
and the kinematic age of the local population of ultracool dwarfs.

We detected the Li line in six of our targets, i.e., about 10\% of our
targets are younger than about 0.5\,Gyr. This result is consistent
with earlier results on the brown dwarf fraction by \citet{Reid02a}.
The mass function index $\alpha$ is perhaps a little higher than
previously thought, but $\alpha < 2$ is probably still valid.

From M7 to mid-L, the Li fraction rises because the cooler stars are
less massive and retain their Li for a longer period. However, we find
a local maximum of the Li fraction at spectral type M9. At this point,
we cannot exclude that this maximum is caused by the full sample being
partially biased towards younger objects that are brighter and hence
easier to observe. A potential explanation for a real Li maximum could
be a situation in which the spectral appearance of lithium brown
dwarfs together with the mass-dependent detectability of the Li line
and the star formation rate results in a maximum of stars with
detectable Li lines at spectral type M9. The details of such a
scenario are not clear, and a statistically more robust sample of
late-M and early-L dwarfs should be investigated before drawing
further conclusions. A necessary condition for this is the
availability of trigonometric parallaxes.

We investigated the space velocities of ultracool dwarfs.
$UVW$-velocities are concentrated around the thin disk, and the
lithium brown dwarfs kinematically cluster around the very center of
the thin disk.  Space velocities of our sample generally follow a
Gaussian distribution so that we can derive a kinematic age for our
sample. The kinematic age of M7--M9.5 dwarfs is 3.1\,Gyr, which
supports earlier results \citep[see][]{Reid02a} that the velocity
dispersion of late-M dwarfs is significantly lower than those measured
for nearby M dwarfs, and that the dispersion is similar to the one in
earlier type emission-line stars.


\acknowledgements

Based on observations collected at the European Southern Observatory,
Paranal, Chile, PIDs 080.D-0140 and 081.D-0190, and observed from the
W.M. Keck Observatory, which is operated as a scientific partnership
among the California Institute of Technology, the University of
California and the National Aeronautics and Space Administration. We
would like to acknowledge the great cultural significance of Mauna Kea
for native Hawaiians and express our gratitude for permission to
observe from atop this mountain. We thank Andreas Seifahrt and Denis
Shulyak for helpful discussions on space velocities and Li abundances.
A.R.  has received research funding from the DFG as an Emmy Noether
fellow (RE 1664/4-1).  G.B. thanks the NSF for grant support through
AST06-06748.

\begin{deluxetable}{llcrrrrrrcc}
  \tablecaption{\label{tab:Sample}Sample targets and their space
    motion. }
  \tablewidth{0pt}
  \tablehead{\colhead{Name} & \colhead{Other Name} & \colhead{SpType} & \colhead{$J$} & \colhead{$d$}& \colhead{$U$} & \colhead{$V$} & \colhead{$W$} & \colhead{$v_{\rm rad}$\tablenotemark{*}} & \colhead{pmRef} & \colhead{Note} \\
    &&&& [pc] & \colhead{[km/s]} & \colhead{[km/s]} & \colhead{[km/s]}
    & \colhead{[km/s]} && } 
  \startdata 
  \input{sample.tex}
 \enddata
 \tablenotetext{*}{The typical uncertainty for $v_{\rm rad}$ is
   1--3\,km\,s$^{-1}$ depending on rotational line broadening.}
 \tablecomments{Li: lithium brown dwarf -- SB1: single lined binary --
   SB2: double lined binary}

 \tablerefs{(a) \citet{RE09}, (b) \citet{Costa05}, (c)
   \citet{Costa06}, (d) \citet{Lepine09}, (e) \citet{Monet92}, (f)
   \citet{Bartlett06}; proper motion usually taken from
   \citet{Faherty09}, except for: (1) \citet{PhanBao01}, (2)
   \citet{PhanBao03}, (3) \cite{Salim03}}

\end{deluxetable}

\clearpage

\begin{deluxetable}{cccc}
  \tablecaption{\label{tab:Li}Li fractions}
  \tablewidth{0pt}
  \tablehead{\colhead{Spectral bin} & \colhead{$N_{\rm Li}$} & \colhead{$N_{\rm tot}$} & \colhead{Li fraction [\%]}}
  \startdata
  M7 -- M9.5 & 6 & 63 & $10^{+4}_{-3}$\\[1ex]
  M8 -- M9.5 & 5 & 39 & $13^{+7}_{-4}$\\[1ex]
  \hline
  \noalign{\smallskip}
  M7/M7.5 & 1 & 24 &   $4^{+9}_{-1}$\\[1ex]
  M8/M8.5 & 2 & 27 &   $7^{+9}_{-2}$\\[1ex]
  M9/M9.5 & 3 & 12 &  $25^{+15}_{-8}$
  \smallskip
  \enddata
\end{deluxetable}

\clearpage

\begin{deluxetable}{lccccc}
  \tablecaption{\label{tab:Ages}Velocity dispersions $\sigma_U,
    \sigma_V$, and $\sigma_W$, total dispersion $\sigma_{\rm tot}$,
    and age constraints from this sample and from the M7--M9.5 sample
    from \citet{Reid02a}.  Velocity dispersions are $|W|$-corrected
    according to \citet{Wielen77}.}  \tablewidth{0pt}
  \tablehead{ & \colhead{$\sigma_U$} & \colhead{$\sigma_V$} & \colhead{$\sigma_W$} & \colhead{$\sigma_{\rm tot}$\tablenotemark{a}} & \colhead{Age}\\
    & [km\,s$^{-1}$] & [km\,s$^{-1}$] & [km\,s$^{-1}$] &
    [km\,s$^{-1}$] & [Gyr]} \startdata
  this work       & 30.7 & 22.6 & 16.0 & 41.3 & 3.1 \\
  \citet{Reid02a} & 33.7 & 17.4 & 17.4 & 41.7 & 3.1
  \enddata
  \tablenotetext{a}{$\sigma_{\rm tot} = (\sigma_U^2 + \sigma_V^2 + \sigma_W^2)^{1/2}$}
\end{deluxetable}

\end{document}

%% file: sample.tex
2MASSJ~$ 0019262+461407$ &                      & M8.0 & 12.60 & 19.50 & $  0$ & $-25$ & $ -5$ & $ -19.5$ &  Li      \\
2MASSJ~$ 0019457+521317$ &                      & M9.0 & 12.79 & 18.70 & $ -6$ & $-38$ & $ -6$ & $ -29.1$ & --       \\
2MASSJ~$ 0024246-015819$ &        BRI 0021-0214 & M9.5 & 11.99 & 11.55 & $ -1$ & $ 12$ & $ -6$ & $  10.4$ & --       \\
2MASSJ~$0024442-270825B$ &      LHS 1070B       & M8.5 &  9.25 &  7.71 & $-10$ & $ 20$ & $ 35$ & $ -35.1$ & SB1      \\
2MASSJ~$0027559+221932A$ &           LP 349-25A & M8.0 & 10.61 & 10.30 & $ -8$ & $-26$ & $  3$ & $ -16.8$ & SB1      \\
2MASSJ~$ 0041353-562112$ &                      & M7.5 & 11.96 & 17.00 & $  0$ & $-12$ & $ -3$ & $   6.8$ &  Li      \\
2MASSJ~$ 0109511-034326$ &            LP 647-13 & M9.0 & 11.69 & 11.10 & $-14$ & $-12$ & $  8$ & $  -6.5$ & --       \\
2MASSJ~$ 0123112-692138$ &                      & M8.0 & 12.32 & 17.20 & $ -2$ & $-13$ & $ -5$ & $  10.9$ &  Li      \\
2MASSJ~$ 0148386-302439$ &                      & M7.5 & 12.30 & 18.40 & $  0$ & $ 11$ & $-12$ & $   9.6$ & --       \\
2MASSJ~$ 0248410-165121$ &      LP 771-21/BR 02 & M8.0 & 12.55 & 16.23 & $ 11$ & $-17$ & $ -7$ & $   4.1$ & --       \\
2MASSJ~$ 0306115-364753$ &                      & M8.5 & 11.69 & 11.30 & $ 32$ & $-21$ & $-11$ & $  10.6$ & --       \\
2MASSJ~$ 0320596+185423$ &           LP 412- 31 & M8.0 & 11.76 & 14.51 & $-45$ & $-19$ & $-22$ & $  44.9$ & --       \\
2MASSJ~$ 0331302-304238$ &           LP 888- 18 & M7.5 & 11.36 & 12.10 & $  8$ & $-26$ & $-18$ & $  23.4$ & --       \\
2MASSJ~$ 0334114-495334$ &                      & M9.0 & 11.38 &  8.20 & $-66$ & $-95$ & $ -8$ & $  71.4$ & --       \\
2MASSJ~$ 0339352-352544$ &      LP 944- 20/ BRI & M9.0 & 10.73 &  4.97 & $-12$ & $ -5$ & $ -2$ & $   7.6$ &  Li      \\
2MASSJ~$ 0351000-005244$ &             LHS 1604 & M7.5 & 11.30 & 14.66 & $ 28$ & $-27$ & $ -6$ & $ -14.7$ & --       \\
2MASSJ~$ 0417374-080000$ &                      & M7.5 & 12.18 & 17.40 & $-39$ & $-40$ & $  3$ & $  38.4$ & --       \\
2MASSJ~$0429184-312356A$ &                      & M7.5 & 10.87 & 11.40 & $-24$ & $-23$ & $-23$ & $  39.6$ & --       \\
2MASSJ~$ 0435161-160657$ &           LP 775- 31 & M7.0 & 10.41 &  8.60 & $-46$ & $-19$ & $-23$ & $  52.5$ & SB2      \\
2MASSJ~$ 0440232-053008$ &           LP 655- 48 & M7.0 & 10.66 &  9.80 & $-29$ & $-14$ & $  0$ & $  27.5$ & --       \\
2MASSJ~$ 0443376+000205$ &       SDSS 0443+0002 & M9.0 & 12.51 & 16.20 & $-11$ & $-11$ & $-10$ & $  17.1$ &  Li      \\
2MASSJ~$ 0517376-334902$ &      DENIS-P J0517-3 & M8.0 & 12.00 & 14.70 & $  1$ & $-48$ & $  3$ & $  31.4$ & --       \\
2MASSJ~$ 0544115-243301$ &                      & M8.0 & 12.53 & 18.70 & $ 37$ & $-52$ & $-17$ & $  20.8$ & --       \\
2MASSJ~$ 0741068+173845$ &             LHS 1937 & M7.0 & 12.01 & 17.90 & $-31$ & $-48$ & $-18$ & $  38.6$ & --       \\
2MASSJ~$ 0752239+161215$ &           LP 423- 31 & M7.0 & 10.88 & 10.50 & $ 24$ & $-11$ & $ -5$ & $ -18.4$ & --       \\
2MASSJ~$ 0818580+233352$ &                      & M7.0 & 12.18 & 19.10 & $ -9$ & $-40$ & $ 28$ & $  33.1$ & --       \\
2MASSJ~$ 0853362-032932$ &             LHS 2065 & M9.0 & 11.21 &  8.53 & $-13$ & $ -9$ & $-17$ & $   6.4$ & --       \\
2MASSJ~$ 1006319-165326$ &           LP 789- 23 & M7.5 & 12.04 & 16.40 & $-31$ & $-19$ & $ 10$ & $  27.8$ & --       \\
2MASSJ~$ 1016347+275149$ &             LHS 2243 & M8.0 & 11.99 & 14.40 & $ -8$ & $-39$ & $ 14$ & $  24.4$ & --       \\
2MASSJ~$ 1024099+181553$ &                      & M8.0 & 12.27 & 16.50 & $-10$ & $ -9$ & $ -2$ & $   6.0$ & --       \\
2MASSJ~$ 1048126-112009$ &                      & M7.0 &  8.86 &  4.50 & $ 28$ & $-16$ & $-14$ & $  -0.2$ & --       \\
2MASSJ~$ 1048147-395606$ &                      & M9.0 &  9.54 &  4.10 & $-11$ & $  2$ & $-30$ & $ -12.9$ & --       \\
2MASSJ~$1121492-131308A$ &           LHS 2397aA & M8.0 & 11.93 & 14.45 & $-25$ & $-36$ & $  9$ & $  31.8$ & SB1      \\
2MASSJ~$ 1124048+380805$ &                      & M8.5 & 12.71 & 19.00 & $ 16$ & $  2$ & $ -9$ & $ -14.0$ & --       \\
2MASSJ~$ 1141440-223215$ &                      & M8.0 & 12.63 & 17.20 & $-23$ & $  5$ & $ 26$ & $   8.4$ & --       \\
2MASSJ~$ 1155429-222458$ &                      & M7.5 & 10.93 &  9.70 & $-14$ & $  0$ & $-16$ & $ -13.1$ & --       \\
2MASSJ~$ 1224522-123835$ &         BR 1222-1221 & M9.0 & 12.57 & 17.06 & $-13$ & $-17$ & $-16$ & $  -5.8$ & --       \\
2MASSJ~$ 1246517+314811$ &             LHS 2632 & M7.5 & 12.23 & 18.10 & $-60$ & $-32$ & $  4$ & $   5.1$ & --       \\
2MASSJ~$ 1253124+403403$ &             LHS 2645 & M7.5 & 12.19 & 17.50 & $ 43$ & $-30$ & $ 15$ & $   3.6$ & --       \\
2MASSJ~$ 1309218-233035$ &               CE 303 & M8.0 & 11.78 & 13.30 & $ 19$ & $-25$ & $ -8$ & $  19.3$ & --       \\
2MASSJ~$ 1332244-044112$ &                      & M7.5 & 12.37 & 18.90 & $ -4$ & $ 13$ & $-13$ & $ -16.9$ & --       \\
2MASSJ~$ 1356414+434258$ &           LP 220- 13 & M7.0 & 11.71 & 15.60 & $-29$ & $-26$ & $-14$ & $ -22.2$ & --       \\
2MASSJ~$ 1403223+300754$ &                      & M8.5 & 12.68 & 18.80 & $-60$ & $-55$ & $-21$ & $ -42.4$ & --       \\
2MASSJ~$ 1411213-211950$ &                      & M9.0 & 12.44 & 15.70 & $ -2$ & $  5$ & $  5$ & $  -0.8$ &  Li      \\
2MASSJ~$ 1438082+640836$ &                      & M9.5 & 12.98 & 18.40 & $ 57$ & $  1$ & $-47$ & $ -44.9$ & --       \\
2MASSJ~$ 1440229+133923$ &      LSPM J1440+1339 & M8.0 & 12.40 & 17.60 & $  7$ & $-30$ & $ -7$ & $  -5.2$ & --       \\
2MASSJ~$ 1456383-280947$ &             LHS 3003 & M7.0 &  9.97 &  6.37 & $ -5$ & $-26$ & $-12$ & $   0.9$ & --       \\
2MASSJ~$ 1507277-200043$ &                      & M7.5 & 11.71 & 14.20 & $  3$ & $  2$ & $ -8$ & $  -2.5$ & --       \\
2MASSJ~$ 1521010+505323$ &                      & M7.5 & 12.01 & 16.10 & $ 17$ & $ -3$ & $  2$ & $   0.9$ & --       \\
2MASSJ~$ 1534570-141848$ &                      & M7.0 & 11.38 & 13.50 & $-85$ & $-44$ & $-19$ & $ -75.5$ & --       \\
2MASSJ~$ 1546054+374946$ &                      & M7.5 & 12.44 & 19.70 & $  1$ & $-21$ & $-18$ & $ -24.9$ & --       \\
2MASSJ~$ 1757154+704201$ &            LP 44-162 & M7.5 & 11.45 & 12.50 & $-18$ & $-15$ & $ -8$ & $ -13.5$ & --       \\
2MASSJ~$ 1835379+325954$ &       LSR J1835+3259 & M8.5 & 10.27 &  5.67 & $ 22$ & $ -1$ & $ -3$ & $   8.5$ & --       \\
2MASSJ~$ 1843221+404021$ &             LHS 3406 & M8.0 & 11.31 & 14.14 & $-42$ & $-14$ & $ 13$ & $ -22.3$ & --       \\
2MASSJ~$ 2037071-113756$ &                      & M8.0 & 12.27 & 16.80 & $-16$ & $-47$ & $  7$ & $ -38.7$ & --       \\
2MASSJ~$ 2206227-204706$ &                      & M8.0 & 12.37 & 18.20 & $  5$ & $  1$ & $ -9$ & $   9.8$ & SB1      \\
2MASSJ~$ 2226443-750342$ &                      & M8.5 & 12.35 & 16.50 & $  5$ & $ -8$ & $-12$ & $  14.7$ & --       \\
2MASSJ~$ 2237325+392239$ &             G 216-7B & M9.5 & 13.34 & 18.89 & $ 21$ & $-64$ & $ -9$ & $ -61.2$ & --       \\
2MASSJ~$ 2306292-050227$ &                      & M8.0 & 11.35 & 11.00 & $-41$ & $-64$ & $ 19$ & $ -56.3$ & --       \\
2MASSJ~$ 2331217-274949$ &                      & M8.5 & 11.65 & 11.60 & $-22$ & $ 36$ & $  3$ & $  -4.2$ & --       \\
2MASSJ~$ 2349489+122438$ &           LP 523- 55 & M8.0 & 12.60 & 19.60 & $  8$ & $-15$ & $-10$ & $  -3.5$ & --       \\
2MASSJ~$ 2351504-253736$ &                      & M8.0 & 12.47 & 18.20 & $-37$ & $ -3$ & $  3$ & $ -10.0$ & --       \\
2MASSJ~$ 2353594-083331$ &                      & M8.5 & 13.03 & 19.10 & $ 49$ & $ 36$ & $-18$ & $  32.7$ & --       \\